\documentclass[10pt]{book}

\usepackage{theorem}
\theoremstyle{plain}{

}

\theoremstyle{plain}{

}
\usepackage{fancyhdr}
\renewcommand{\sectionmark}[1]{\markright{{\sf   {\sl Volume $\bullet$}, $\bullet$. }}{}}
\usepackage[Lenny]{fncychapmsel}
\usepackage{amssymb,amsmath,amsfonts}
\usepackage[latin1]{inputenc}
\usepackage{enumerate}
\usepackage[greek,spanish,french,english]{babel}
\usepackage[mathscr]{eucal}
\usepackage{color}
\usepackage{graphicx}
\usepackage{booktabs}
\usepackage{minitoc}
\def\abst#1{
\begin{center}
\begin{minipage}{17cm} \it
\hrule
\begin{center}
$\vspace{0.2cm}$\\
  {\bf Abstract}\\
\end{center}
\begin{center}
\begin{minipage}{16cm}
  #1
\end{minipage}
\end{center}
\hrule

\end{minipage}
\end{center}}
\definecolor{Azulon}{rgb}{0.65,0.75,1}
\newcommand\logoMSEL{
\begin{minipage}{4cm}
\resizebox{4cm}{!}{\textcolor{blue}{ \fontsize{14}{16}\usefont{OT1}{pag}{m}{n}\selectfont \bf \color{Azulon}{MSEL}}}
\resizebox{4cm}{2mm}{\sc \hspace{2.5cm} \rotatebox{90}{\hspace{-1.8cm} Modelling} in Science Education and Learning}
\end{minipage}}

\newenvironment{keyword}{{\sf Keywords: }}%
{\endtrivlist\medskip}

\newenvironment{clave}{{\sf Palabras clave: }}%
{\endtrivlist\medskip}

{\endtrivlist\medskip}

{\endtrivlist\medskip}

\usepackage{hyperref}
\hypersetup{pdfpagemode=UseOutlines,colorlinks=true,citecolor=blue}
\usepackage{apacite}

\makeatletter
   \def\cleardoublepage{\clearpage\if@twoside \ifodd\c@page\else
  \vspace*{\fill}
     \thispagestyle{empty}
    \newpage
    \if@twocolumn\hbox{}\newpage\fi\fi\fi}
\makeatother

\def\vol{$\bullet$}
\def\anyo{$\bullet$}

\usepackage[T1]{fontenc}

\usepackage[spanish]{babel}

\begin{document}

\renewcommand{\bibname}{R\MakeLowercase{eferences}}
\renewcommand{\tablename}{Table}
\renewcommand{\figurename}{Figure}

\renewcommand{\thetable}{\arabic{table}}
\renewcommand{\theequation}{\arabic{equation}}
\renewcommand{\thefigure}{\arabic{figure}}


\normalsize
\chapter[Short title \\
\color{black} {\sc Monteiro M., Stari C., Marti A.C.}]{Mobile-device
  sensors: an innovative tool in the teaching of physical
  sciences.\\ Los sensores de los dispositivos móviles: una
  herramienta innovadora en la enseñanza de las ciencias físicas}
\minitoc


\renewcommand\author1{
\begin{minipage}{.4\textwidth}
{\bf  Martín Monteiro} \\ 
{\sc Universidad ORT Uruguay, Uruguay.} \\ 
{\href{mailto:fisica.martin@gmail.com}{fisica.martin@gmail.com}}
 \end{minipage}}
 \hfill \renewcommand\author2{\begin{minipage}{.4\textwidth} {\bf
       Cecilia Stari} \\ {\sc Facultad de Ingeniería, Universidad de
       la República, Montevideo, Uruguay.}
     \\ {\href{mailto:cstari@fing.edu.uy}{cstari@fing.edu.uy}}
\end{minipage}}
 \hfill \renewcommand\author3{\begin{minipage}{.4\textwidth} {\bf
       Arturo C. Marti} \\ {\sc Facultad de Ciencias, Universidad de
       la República, Montevideo, Uruguay.}
     \\ {\href{mailto:marti@fisica.edu.uy}{marti@fisica.edu.uy}}
\end{minipage}}


\abst{\large We show how builtin sensors in mobile devices can be used
  as portable laboratories at the service of teaching experimental
  sciences, especially physics, in the last years of high school and
  first years of university. We describe experiments that previously
  required expensive apparatus or were not feasible in teaching
  laboratories. Finally, we discuss some perspectives about the use of
  sensors in the physics teaching.  \\[2mm] Mostramos como los sensores
  incorporados en dispositivos móviles pueden ser utilizados como
  laboratorios portátiles al servicio de la enseñanza de las ciencias
  experimentales, especialmente de la física, en los últimos años de
  la educación media y los primeros de la universitaria. Describimos
  experimentos que antes requerían costosos aparatos o que no eran
  factibles en laboratorios de enseñanza. Finalmente, discutimos
  algunas perspectivas del uso de los sensores en la enseñanza de la
  física.  }

\noindent \keyword{mobile devices, sensors, smartphones, physics teaching

\noindent \clave{dispositivos móviles, sensores, teléfonos inteligentes}
\newpage
\large

\section{Introducción} 
Los teléfonos inteligentes se popularizaron notablemente en los
últimos años. Actualmente su uso trasciende considerablemente el
propósito original de hablar con un interlocutor distante. De hecho,
cada día es más frecuente utilizar dispositivos móviles como relojes,
cámaras, agendas, reproductores de música o GPS. Resulta destacable el
hábito presente en todos los estratos sociales, pero especialmente
llamativo entre los jóvenes, de llevar sus teléfonos inteligentes todo
el tiempo y a todas partes. El conocido portal statistia.com estima en
base a fuentes confiables, en todo el mundo, más de 1.500 millones
celulares inteligentes fueron vendidos a usuarios finales durante el
año 2021.

Una característica de los teléfonos inteligentes no siempre es
conocida por los usuarios esla incorporación de varios sensores, entre
ellos el acelerómetro, el sensor de velocidad angular, el
magnetómetro, el sensor de proximidad o el de presión. Estos sensores
facilitan la experiencia de los usuarios en diversos sentidos, por
ejemplo, desactivando la pantalla táctil para evitar acciones no
deseadas, regulando el brillo de la pantalla para adaptarse a las
condiciones de iluminación ambiente, o habilitando comando gestuales
como sacudir el dispositivo para realizar rápidamente algunas
acciones. A pesar que entre el propósito de los fabricantes no es
hacer experimentos de física, es posible reinventar el uso de estos
dispositivos y ponerlos al servicio de la enseñanza de las ciencias
físicas.

Los sensores de los dispositivos móviles permiten experimentar
obteniendo mediciones reales de diversos fenómenos físicos. Es posible
medir aceleración, velocidad angular, campo magnético, presión,
sonido, entre muchas otras magnitudes, especialmente en laboratorios
de secundaria o pregrado. Sin embargo, gracias a su portabilidad los
experimentos con teléfonos inteligentes se pueden realizar fácilmente
en ubicaciones no tradicionales como parques infantiles, instalaciones
deportivas o medios de transporte (véase la columna iPhysicsLab
publicada por The Physics Teacher o también la reciente revisión,
\cite{monteiro2022resource}). En términos generales, los dispositivos
móviles permiten dos formas de trabajo, una, en el propio dispositivo
y otra como data logger, transfiriendo la información a un ordenador
tradicional o trabajando en la nube. En la Tabla 1 se enumeran los
sensores más frecuentes en la actualidad junto con sus principales
características.

\begin{table}[h!]
\caption{Sensores más frecuentes en los dispositivos móviles actuales y sus características. 
Se entiende por seudosensor un dispositivo que incorpora información de varios sensores y 
por medio de un procesamiento provee una nueva magnitud.} 
\begin{center}
\begin{tabular}{|c|c|c|c|}   \hline
Sensor &Propósito básico & Cantidad & Magnitud física \\ \hline  \hline
Acelerómetro&  Rotar pantalla, detectar movimiento&  Vector&  Aceleración  \\  \hline 
Giroscopio & Determinar orientación  dispositivo &Vector& Velocidad angular  \\  \hline
Micrófono & Comunicación & escalar& Sonido  \\  \hline 
Cámara & Obtener vídeos, imágenes & Matricial&  Mapa de imágenes  \\  \hline 
Proxímetro&  Desactivar la pantalla & Booleana& Objeto próximo detectado  \\   \hline
Luxómetro &Controlar la pantalla& Escalar & Luz ambiente  \\  \hline 
Presión & Localización y determinación de  & Escalar &Presión atmosférica  \\ 
 &  altura en interiores y exteriores & & \\ \hline 
Orientación (seudosensor)& Determinar orientación &3 ángulos &Alabeo, guiñada y cabeceo  \\   \hline
GPS (receptor) & Geolocalización& Geolocalización &Coordenadas geográficas  \\   \hline
Magnetómetro &Orientación& Vector &Campo magnético local  \\   \hline
\end{tabular}
\end{center}
\label{tab1}
\end{table}

\section{Objetivos} 
En los últimos años se han propuesto numerosos experimentos de física
basados en los dispositivos móviles. El número de estos trabajos
creció en forma explosiva desde la aparición de los primeros teléfonos
inteligentes en la década del 2000. En la figura 1 recapitulamos la
cantidad de trabajos publicados, de acuerdo al conocido portal Scopus,
en revistas de ciencias físicas e ingenierías que contienen términos
relacionados con este trabajo en el título, resumen y palabras
claves. Estos experimentos abarcan todas las áreas de la física, desde
mecánica clásica hasta física moderna.

\begin{figure}[h!]
   \hfil
\includegraphics[scale=.6]{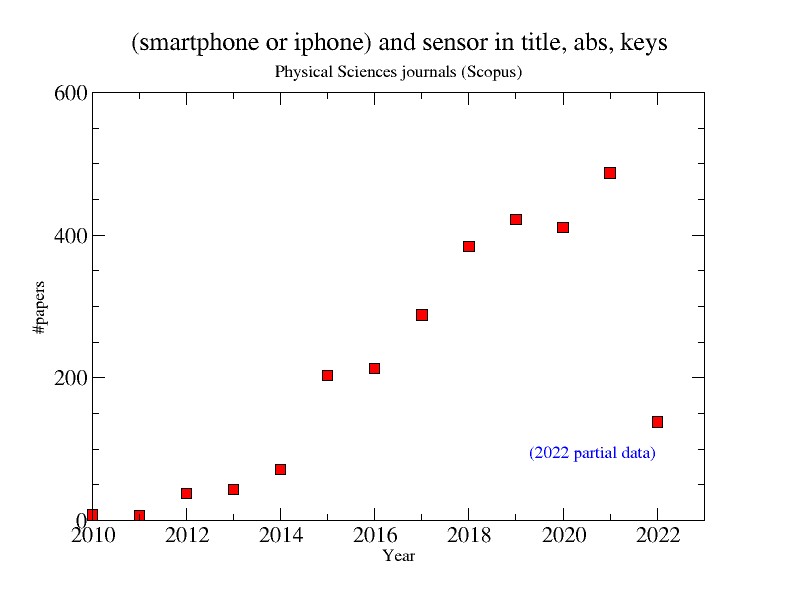}
    \caption{Evolución anual del número de artículos publicados en
      revistas de ciencias físicas y de ingenierías vinculadas con
      dispositivos móviles y sensores de acuerdo al portal
      Scopus.} \label{Fig:1}
\end{figure}

El objetivo de este trabajo es mostrar cómo es posible utilizar los
sensores de los dispositivos móviles en la experimentación de
física. La utilización de estos dispositivos cobró especial
importancia durante la pandemia de covid que obligó a replantear las
metodologías en un ámbito global \cite{o2021guide}. Estas
transformaciones no implican solamente un cambio en los instrumentos
sino que apunta a diseñar y promover estrategias de enseñanza y de
aprendizaje activo. Buscamos cambiar el rol de la enseñanza y otorgar
un papel fundamental al estudiante, dejando de ser un receptor pasivo
del conocimiento e involucrándose en las actividades en el aula.

\section{Metodología}

Las nuevas tecnologías abren la puerta a nuevas oportunidades y
proporcionan un marco ideal para propiciar el trabajo en equipo en un
contexto de inclusión e igualdad. Éstas, sin embargo, por sí solas no
alcanzan para transformar la educación. En palabras de la OCDE \textit{la
tecnología puede amplificar las buenas prácticas docentes pero ni la
mejor tecnología puede solucionar la falta de buena
docencia}. Siguiendo esta filosofía, propiciamos un salto cualitativo
en la enseñanza de la Física y de las ciencias naturales gracias a la
integración de los sensores de los dispositivos móviles. Apuntamos
entonces a contribuir a mejorar el sistema educativo y a prepararlo
para los cambios tecnológicos que se avecinan, en particular, con la
ayuda de las tecnologías digitales y el pensamiento computacional. Nos
basamos en el diseño de experiencias utilizando tecnologías digitales
y la evaluación de su efecto en la enseñanza de las ciencias naturales
y de la Física en particular.

\selectlanguage{spanish}

\begin{figure}[h!]
   
\hfil
\includegraphics[scale=.19]{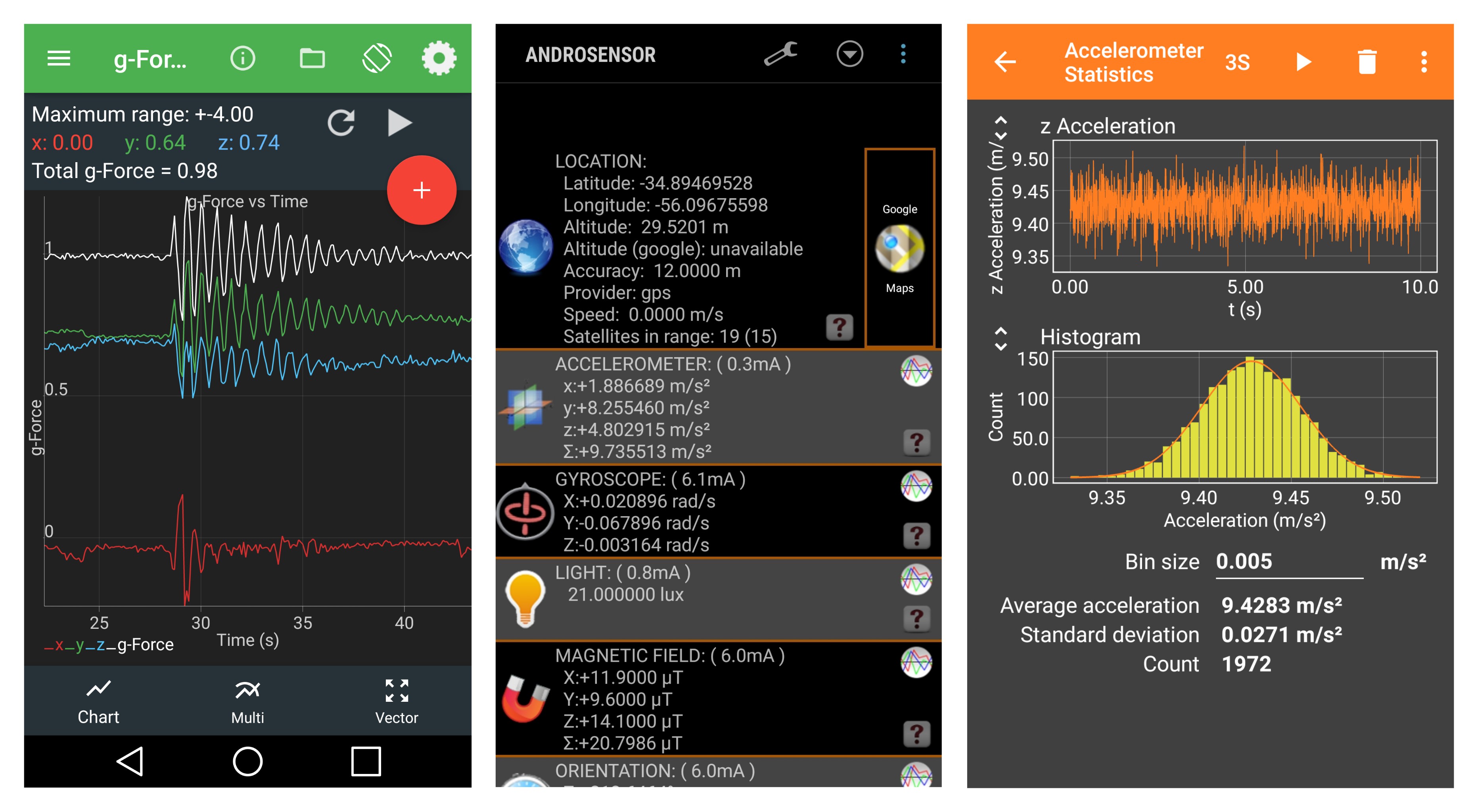}
    \caption{Captura de pantalla de algunas aplicaciones muy utilizadas en la actualidad.} \label{Fig:2}
\end{figure}

El uso de los sensores requiere de paquetes específicos de software,
usualmente referidos como \textit{apps} para poder acceder a los datos
relevados. En las tiendas de apps de cada sistema operativo se cuenta
con numerosas opciones. En la figura 2 mencionamos tres de las mejor
adaptadas para realizar experimentos de física: Physics Toolbox Suite,
Phyphox y Androsensor. Cada una de ellas tienen sus propias
funcionalidades como acceder en forma remota, programación basada en
navegadores o análisis de datos en la propia aplicación. Otros
aspectos importantes son la facilidad para realizar diversas medidas,
utilizar varios sensores en forma simultánea y configurar opciones
como la frecuencia de muestreo, las unidades o el intervalo de
medición. Otras características útiles incluyen la posibilidad de
compartir experimentos en forma colaborativa con otros usuarios \cite{staacks_2018}.
Gracias a la gran potencia de cálculo es posible realizar diversas operaciones, algunas sencillas como ajustes lineales
o no lineales o también realizar transformadas rápidas de Fourier en
tiempo real.

\section{Resultados y Discusión}

Un conjunto muy diverso de experimentos puede ser realizado siguiendo
esta metodología. A continuación planteamos una selección de aquellos
que consideramos más ilustrativos. Todo resultado de medición debe ser
expresado con la incertidumbre correspondiente. El estudio de la
teoría de errores es un elemento fundamental en la formación de
futuros científicos e ingenieros. En las prácticas usuales durante
mucho tiempo se estudió este tema tomando medidas reiteradas de un
fenómeno, por ejemplo arrojando una bolita y midiendo la distancia
para luego construir gráficas e histogramas. Los sensores permiten
transformar este tipo de experimentos tomando cientos o miles de
medidas en pocos segundos \cite{monteiro2021using} y luego estudiar conceptos
tales como la desviación estándar, el error estándar, distribuciones
normales y criterios estadísticos.

La experimentación con juguetes también es un punto alto de los
dispositivos móviles \cite{Salinas2020dynamics}. En este trabajo mostramos cómo es
posible estudiar la dinámica clásica de un tradicional juguete, el
yoyo, utilizando el giroscopio de un teléfono móvil. Gracias a este
sensor se obtienen medidas muy complicadas de obtener con otros
instrumentos. Otro ejemplo de la versatilidad de estos dispositivos es la posiblidad de
analizar los prinicipios físicos del vuelo de drones \cite{monteiro2022simple} o de los tubos
de simples aspiradoras \cite{monteiro2022home}.

Como mencionamos antes, otra ventaja es la posibilidad
de medir en sitios no tradicionales. En la referencia \cite{monteiro2020magnetic},
planteamos un experimento donde medimos el campo magnético generado
por trenes eléctricos. Esta medida es complementada con datos
obtenidos en internet del campo magnético terrestre local, de la
orientación geográfica y de las características técnicas de las líneas
férreas.  El primer iPhone, lanzado en 2007, incorporó varios avances
tecnológicos que ya estaban presentes en otros dispositivos, sin
embargo, tuvo la gran virtud de reunir todas estas características en
un solo dispositivo portátil y popular. Se puede hacer una observación
similar sobre el uso de teléfonos inteligentes en un laboratorio de
física, todos los ingredientes, como el micrófono, la cámara o los
sensores, ya son conocidos y están disponibles en otros dispositivos
individuales, sin embargo, los teléfonos inteligentes presentan la
ventaja de recopilar un gran cantidad de herramientas listas para usar
en un solo dispositivo.

\section{Conclusiones}

Como inferimos de las consideraciones anteriores los experimentos
basados en sensores de dispositivos móviles constituyen una
herramienta de innovación en la enseñanza de las ciencias físicas. Las
ventajas son numerosas por un lado son dispositivos de uso corriente
por parte de los estudiantes de todos los niveles que no los perciben
como elementos extraños sino como objetos corrientes. Su amplia
disponibilidad conlleva que el precio es relativamente accesible en
comparación con otros sensores de fabricación específica para
laboratorios o centros educativos. Además, el dispositivo móvil ya
está disponible y no significa un gasto adicional. Por otro lado,
estos integran varios sensores y es posible usarlos simultáneamente en
un mismo experimento sin necesidad de otros instrumentos de
comunicación \cite{monteiro2019physics}.

La cantidad de dispositivos vendidos en todo el mundo continua
aumentando. Asimismo, vienen cada vez mejor equipados y con más
sensores. Cabe preguntarse si el uso de sensores en dispositivos
móviles representa un avance significativo o solo marginal. La misma
pregunta se podría hacer sobre el uso de los smartphones en otros
ámbitos, y creemos que las respuestas son esencialmente las mismas:
permiten acceder a un conjunto amplio de recursos en un único
dispositivo, a un precio razonable, de tamaño práctico y facilidades
de uso.  Son muchas las incógnitas que se presentan para los próximos
años. La aparición de los sensores de los dispositivos móviles es
relativamente reciente. Por un lado, no sabemos si los fabricantes
seguirán incluyendo nuevos sensores, es decir sensores que no han
estado disponibles hasta ahora en dispositivos móviles. Por otro lado,
tampoco podemos decir si la capacidad de cálculo está alcanzando un
ritmo de crecimiento más lento o seguirá aumentando al mismo ritmo que
en el actual.
 
Quizás los avances futuros se centren en mejoras relacionadas con una
mayor facilidad de uso del software y el procesamiento de datos. La
investigación en enseñanza de la física también tiene mucho que decir,
evaluando si el cambio futuro será cualitativo o meramente marginal y
si se observan mejoras en los resultados de aprendizaje. En todo caso,
podemos afirmar que estos dispositivos y sensores llegaron para
permanecer por un tiempo considerable. Como ocurre con todas las
nuevas herramientas o tecnologías que surgen, la mejor actitud parece
ser analizar sus aportes y utilizarlos de manera crítica y reflexiva.

\bibliographystyle{apacite}
\bibliography{/home/arturo/Dropbox/bibtex/mybib}

\begin{thebibliography}{}

\bibitem [\protect \citeauthoryear {%
Monteiro%
\ \BBA {} Mart{\'\i}%
}{%
Monteiro%
\ \BBA {} Mart{\'\i}%
}{%
{\protect \APACyear {2022}}%
}]{%
monteiro2022resource}
\APACinsertmetastar {%
monteiro2022resource}%
\begin{APACrefauthors}%
Monteiro, M.%
\BCBT {}\ \BBA {} Mart{\'\i}, A\BPBI C.%
\end{APACrefauthors}%
\unskip\
\newblock
\APACrefYearMonthDay{2022}{}{}.
\newblock
{\BBOQ}\APACrefatitle {Resource Letter MDS-1: Mobile devices and sensors for
  physics teaching} {Resource letter mds-1: Mobile devices and sensors for
  physics teaching}.{\BBCQ}
\newblock
\APACjournalVolNumPages{American Journal of Physics}{90}{5}{328--343}.
\PrintBackRefs{\CurrentBib}

\bibitem [\protect \citeauthoryear {%
Monteiro%
, Organtini%
\BCBL {}\ \BBA {} Mart{\'\i}%
}{%
Monteiro%
\ \protect \BOthers {.}}{%
{\protect \APACyear {2020}}%
}]{%
monteiro2020magnetic}
\APACinsertmetastar {%
monteiro2020magnetic}%
\begin{APACrefauthors}%
Monteiro, M.%
, Organtini, G.%
\BCBL {}\ \BBA {} Mart{\'\i}, A\BPBI C.%
\end{APACrefauthors}%
\unskip\
\newblock
\APACrefYearMonthDay{2020}{}{}.
\newblock
{\BBOQ}\APACrefatitle {Magnetic fields produced by electric railways} {Magnetic
  fields produced by electric railways}.{\BBCQ}
\newblock
\APACjournalVolNumPages{The Physics Teacher}{58}{8}{600--601}.
\PrintBackRefs{\CurrentBib}

\bibitem [\protect \citeauthoryear {%
Monteiro%
, Stari%
, Cabeza%
\BCBL {}\ \BBA {} Marti%
}{%
Monteiro%
\ \protect \BOthers {.}}{%
{\protect \APACyear {2019}}%
}]{%
monteiro2019physics}
\APACinsertmetastar {%
monteiro2019physics}%
\begin{APACrefauthors}%
Monteiro, M.%
, Stari, C.%
, Cabeza, C.%
\BCBL {}\ \BBA {} Marti, A\BPBI C.%
\end{APACrefauthors}%
\unskip\
\newblock
\APACrefYearMonthDay{2019}{}{}.
\newblock
{\BBOQ}\APACrefatitle {Physics experiments using simultaneously more than one
  smartphone sensors} {Physics experiments using simultaneously more than one
  smartphone sensors}.{\BBCQ}
\newblock
\BIn{} \APACrefbtitle {Journal of Physics: Conference Series} {Journal of
  physics: Conference series}\ (\BVOL\ 1287, \BPG~012058).
\PrintBackRefs{\CurrentBib}

\bibitem [\protect \citeauthoryear {%
Monteiro%
, Stari%
, Cabeza%
\BCBL {}\ \BBA {} Mart{\'\i}%
}{%
Monteiro%
\ \protect \BOthers {.}}{%
{\protect \APACyear {2021}}%
}]{%
monteiro2021using}
\APACinsertmetastar {%
monteiro2021using}%
\begin{APACrefauthors}%
Monteiro, M.%
, Stari, C.%
, Cabeza, C.%
\BCBL {}\ \BBA {} Mart{\'\i}, A\BPBI C.%
\end{APACrefauthors}%
\unskip\
\newblock
\APACrefYearMonthDay{2021}{}{}.
\newblock
{\BBOQ}\APACrefatitle {Using mobile-device sensors to teach students error
  analysis} {Using mobile-device sensors to teach students error
  analysis}.{\BBCQ}
\newblock
\APACjournalVolNumPages{American Journal of Physics}{89}{5}{477--481}.
\PrintBackRefs{\CurrentBib}

\bibitem [\protect \citeauthoryear {%
Monteiro%
, Stari%
, Cabeza%
\BCBL {}\ \BBA {} Mart{\'\i}%
}{%
Monteiro%
, Stari%
, Cabeza%
\BCBL {}\ \BBA {} Mart{\'\i}%
}{%
{\protect \APACyear {2022}}%
}]{%
monteiro2022simple}
\APACinsertmetastar {%
monteiro2022simple}%
\begin{APACrefauthors}%
Monteiro, M.%
, Stari, C.%
, Cabeza, C.%
\BCBL {}\ \BBA {} Mart{\'\i}, A\BPBI C.%
\end{APACrefauthors}%
\unskip\
\newblock
\APACrefYearMonthDay{2022}{}{}.
\newblock
{\BBOQ}\APACrefatitle {Simple physics behind the flight of a drone} {Simple
  physics behind the flight of a drone}.{\BBCQ}
\newblock
\APACjournalVolNumPages{Physics Education}{57}{2}{025029}.
\PrintBackRefs{\CurrentBib}

\bibitem [\protect \citeauthoryear {%
Monteiro%
, Stari%
\BCBL {}\ \BBA {} Mart{\'\i}%
}{%
Monteiro%
, Stari%
\BCBL {}\ \BBA {} Mart{\'\i}%
}{%
{\protect \APACyear {2022}}%
}]{%
monteiro2022home}
\APACinsertmetastar {%
monteiro2022home}%
\begin{APACrefauthors}%
Monteiro, M.%
, Stari, C.%
\BCBL {}\ \BBA {} Mart{\'\i}, A\BPBI C.%
\end{APACrefauthors}%
\unskip\
\newblock
\APACrefYearMonthDay{2022}{}{}.
\newblock
{\BBOQ}\APACrefatitle {A home-lab experiment: resonance and sound speed using
  telescopic vacuum cleaner pipes} {A home-lab experiment: resonance and sound
  speed using telescopic vacuum cleaner pipes}.{\BBCQ}
\newblock
\APACjournalVolNumPages{Physics Education}{58}{1}{013003}.
\PrintBackRefs{\CurrentBib}

\bibitem [\protect \citeauthoryear {%
O'Brien%
}{%
O'Brien%
}{%
{\protect \APACyear {2021}}%
}]{%
o2021guide}
\APACinsertmetastar {%
o2021guide}%
\begin{APACrefauthors}%
O'Brien, D\BPBI J.%
\end{APACrefauthors}%
\unskip\
\newblock
\APACrefYearMonthDay{2021}{}{}.
\newblock
{\BBOQ}\APACrefatitle {A guide for incorporating e-teaching of physics in a
  post-COVID world} {A guide for incorporating e-teaching of physics in a
  post-covid world}.{\BBCQ}
\newblock
\APACjournalVolNumPages{American Journal of Physics}{89}{4}{403--412}.
\PrintBackRefs{\CurrentBib}

\bibitem [\protect \citeauthoryear {%
Salinas%
, Monteiro%
, Mart{\'\i}%
\BCBL {}\ \BBA {} Monsoriu%
}{%
Salinas%
\ \protect \BOthers {.}}{%
{\protect \APACyear {2020}}%
}]{%
Salinas2020dynamics}
\APACinsertmetastar {%
Salinas2020dynamics}%
\begin{APACrefauthors}%
Salinas, I.%
, Monteiro, M.%
, Mart{\'\i}, A\BPBI C.%
\BCBL {}\ \BBA {} Monsoriu, J\BPBI A.%
\end{APACrefauthors}%
\unskip\
\newblock
\APACrefYearMonthDay{2020}{}{}.
\newblock
{\BBOQ}\APACrefatitle {Analyzing the Dynamics of a Yo-Yo Using a Smartphone
  Gyroscope Sensor} {Analyzing the dynamics of a yo-yo using a smartphone
  gyroscope sensor}.{\BBCQ}
\newblock
\APACjournalVolNumPages{The Physics Teacher}{58}{8}{569--571}.
\PrintBackRefs{\CurrentBib}

\bibitem [\protect \citeauthoryear {%
Staacks%
, H\"{u}tz%
, Heinke%
\BCBL {}\ \BBA {} Stampfer%
}{%
Staacks%
\ \protect \BOthers {.}}{%
{\protect \APACyear {2018}}%
}]{%
staacks_2018}
\APACinsertmetastar {%
staacks_2018}%
\begin{APACrefauthors}%
Staacks, S.%
, H\"{u}tz, S.%
, Heinke, H.%
\BCBL {}\ \BBA {} Stampfer, C.%
\end{APACrefauthors}%
\unskip\
\newblock
\APACrefYearMonthDay{2018}{may}{}.
\newblock
{\BBOQ}\APACrefatitle {Advanced tools for smartphone-based experiments:
  phyphox} {Advanced tools for smartphone-based experiments: phyphox}.{\BBCQ}
\newblock
\APACjournalVolNumPages{Physics Education}{53}{4}{045009}.
\newblock
\begin{APACrefDOI} \doi{10.1088/1361-6552/aac05e} \end{APACrefDOI}
\PrintBackRefs{\CurrentBib}

\end{thebibliography}

\end{document}